# Computational strategies for cross-species knowledge transfer

This manuscript was automatically generated on November 4, 2025.

## Authors

- **Hao Yuan**
  0000-0002-8848-1595 · yhadevol · yhbioinfo.bsky.social
  Genetics and Genome Science Program; Ecology, Evolution, and Behavior Program, Michigan State University

- **Christopher A. Mancuso**
  0000-0003-3081-2758 · ChristopherMancuso · christopher-mancuso-97b638133
  Department of Biostatistics & Informatics, University of Colorado Anschutz Medical Campus

- **Kayla Johnson**
  0000-0002-0889-5705 · kaylajohnson · kaylajohnson22
  Department of Biomedical Informatics, University of Colorado Anschutz Medical Campus

- **Ingo Braasch** ✉
  0000-0003-4766-611X · fishevodevogeno.bsky.social
  Department of Integrative Biology; Genetics and Genome Science Program; Ecology, Evolution, and Behavior Program, Michigan State University

- **Arjun Krishnan** ✉
  0000-0002-7980-4110 · arjunkrish · compbiologist.bsky.social
  Department of Biomedical Informatics, University of Colorado Anschutz Medical Campus

✉ — Correspondence possible via GitHub Issues or email to Ingo Braasch <braasch@msu.edu>, Arjun Krishnan <arjun.krishnan@cuanschutz.edu>.

## Abstract

Research organisms provide invaluable insights into human biology and diseases, serving as essential tools for functional experiments, disease modeling, and drug testing. However, evolutionary divergence between humans and research organisms hinders effective knowledge transfer across species. Here, we review state-of-the-art methods for computationally transferring knowledge across species, primarily focusing on methods that utilize transcriptome data and/or molecular networks. Our review addresses four key areas: (1) transferring disease and gene annotation knowledge across species, (2) identifying functionally equivalent molecular components, (3) inferring equivalent perturbed genes or gene sets, and (4) identifying equivalent cell types. We conclude with an outlook on future directions and several key challenges that remain in cross-species knowledge transfer, including introducing the concept of “agnology” to describe functional equivalence of biological entities, regardless of their evolutionary origins. This concept is becoming pervasive in integrative data-driven models where evolutionary origins of functions can remain unresolved.

# Introduction

The use of research organisms, also known as model species or model systems, is essential to biomedical research[1]. Leveraging their resemblance in anatomy, physiology, behavior and genetics to corresponding human conditions, research organisms help scientists investigate a wide range of biomedical phenomena and therapeutic treatments before they are applied to humans. For instance, the zebrafish (Danio rerio) is a well-established vertebrate research organism that has external and fast development in large offspring numbers, transparent embryos, and allows for easy genetic manipulation and drug administration through compound exposure[2].

There are two main reasons that research organisms are critical to studying human phenotypes, processes, and diseases. First, it is most often unethical to study biomedical processes or apply novel therapeutic treatments directly in humans. Although there are many unanswered questions on the ethics of using research organisms[3–5], the knowledge derived from these organisms has helped save countless human lives[2,6]. Secondly, genetic studies in humans typically have very high variability due to confounding effects such as population genetics and environmental factors[7–10], and these confounders can be controlled much more under defined laboratory conditions with research organisms.

However, using research organism data to explore human biology presents its own challenges. Evolutionary divergence often gives rise to similar yet different underlying biological processes between human and research organisms[11], thereby impeding the transfer of knowledge across species. Orthologs, i.e., homologous genes from different species that evolved from a shared ancestral gene by speciation, have taken a privileged role in transferring functional annotations across species. Nevertheless, orthologs may have experienced significant, lineage-specific functional changes across species[12–17]. Additionally, the complex genetics underlying phenotypes/processes shared across species may differ. This is because interactions among genes responsible for a phenotype could be rewired during the evolutionary process, potentially encompassing some de novo genes[18]. Consequently, the effective and precise transfer of knowledge considering de novo genes remains challenging. Finally, since no single research organism can fully recapitulate the entirety of a complex biological condition in humans[1], how do we predict which research organisms capture the different facets of the human biology of interest?

To address these issues, researchers have begun developing computational models that can robustly transfer knowledge between a variety of research organisms and humans to identify functionally similar genes or groups of genes across species regardless of their evolutionary origin. Recent discoveries of such cross-species gene pairs were evidenced by large-scale computational competitions such as sbvIMPROVER Species Translation Challenge[19] and the Critical Assessment of Protein Function Annotation Challenge (CAFA)[20–23].

In this review, we present a suite of cross-species knowledge transfer approaches with a significantly broader scope than previous reviews[24–27]. We comprehensively lay out the landscape of recent and state-of-the-art data-driven strategies, including those that leverage AI and machine learning, to answer four classes of important questions that frequently arise when using research organisms to study biomedical questions and translating findings to humans (Figure 1). **(1)** How to predict disease-gene or function-gene relationships across species (Figure 1a)? **(2)** How to identify functionally equivalent molecular components across species (Figure 1b)? **(3)** How to infer perturbed molecular profiles across species (Figure 1c)? **(4)** How to map equivalent cell types and cell states across species (Figure 1d)?

Instead of discussing methods through an algorithmic lens, we are taking a question-first perspective to provide inspiration and background both for computational researchers interested in developing new methods in this area and for experimental/wet-lab researchers interested in finding and using the best methods in this area. Although we have separated the methods into four sections, many of the approaches described across these sections share similarities in algorithmic design, data types, and output formats. Detailed information about the methods underlying each of the main sections can be found in **Supplementary File 1**. In addition, we have curated numerous datasets and databases used in cross-species methodologies, summarized in **Supplementary File 2** and **Supplementary Note 1**. These resources will enable researchers to test and improve upon existing computational methods in the field.

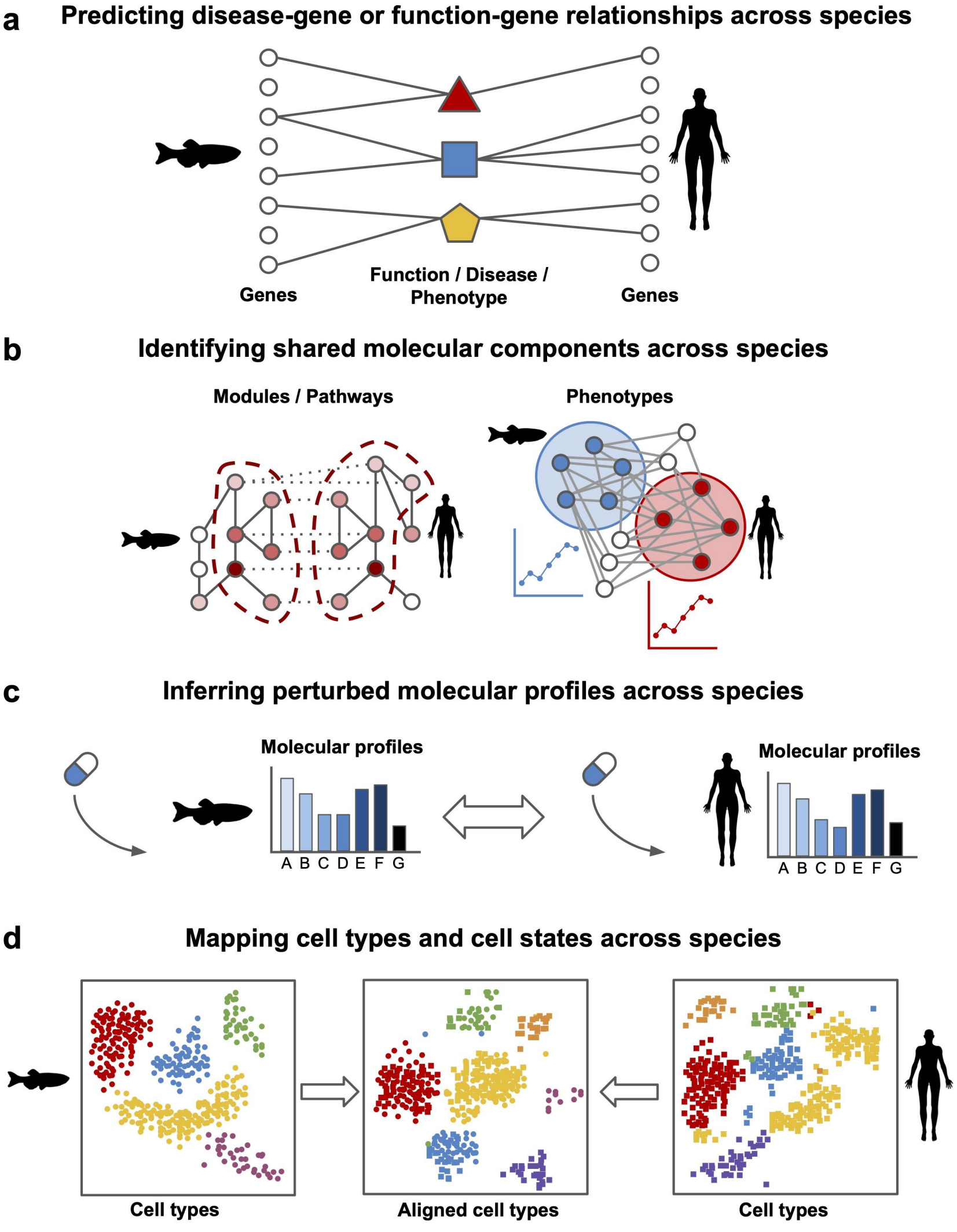

**Figure 1: Schematic diagram of each section of this review article. (a)** How to predict disease-gene or function-gene relationships across species? Diagram depicts associating genes in each species to specific functions, diseases, and phenotypes. **(b)** How to identify functionally equivalent molecular components across species? Diagram depicts finding the most equivalent gene, pathway/expression module, or phenotype between species. **(c)** How to infer perturbed molecular profiles across species? Diagram depicts gene expression in each species as a result of taking a particular perturbation like drug. **(d)** How to map equivalent cell types and cell states across species? Diagram depicts alignment of cell types across species.

# How to predict disease-gene or function-gene relationships across species

Comprehensive knowledge of the roles genes play in molecular functions, phenotypes, and diseases is fundamental to our understanding of the molecular underpinnings of biological, physiological, and pathological processes. However, only a limited number of genes in the human genome have been experimentally characterized. For example, ~33% of protein-coding genes have no functional annotation to any Gene Ontology term[28]. Knowledge of gene-function, gene-phenotype, and gene-disease relationships is significantly richer, though far from complete, in research organisms due to the availability of a variety of experimental techniques such as gene editing, knock-in, and knockout experiments[29,30]. How do we leverage this knowledge available across species to close massive annotation gaps, especially in light of the potential functional divergence of homologous genes and the presence of gene duplications, de novo genes, and gene losses? Moreover, how do we use existing knowledge in human and traditional research organisms (e.g. mouse, frog, zebrafish, fly, worm) to fill annotation gaps in non-traditional research organisms (e.g., dog[31], python[32], gar[33], planaria[34]) that still lack sufficient data. These needs have spurred the development of a number of computational methods for predicting disease-gene or function-gene relationships across multiple species (Figure 2).

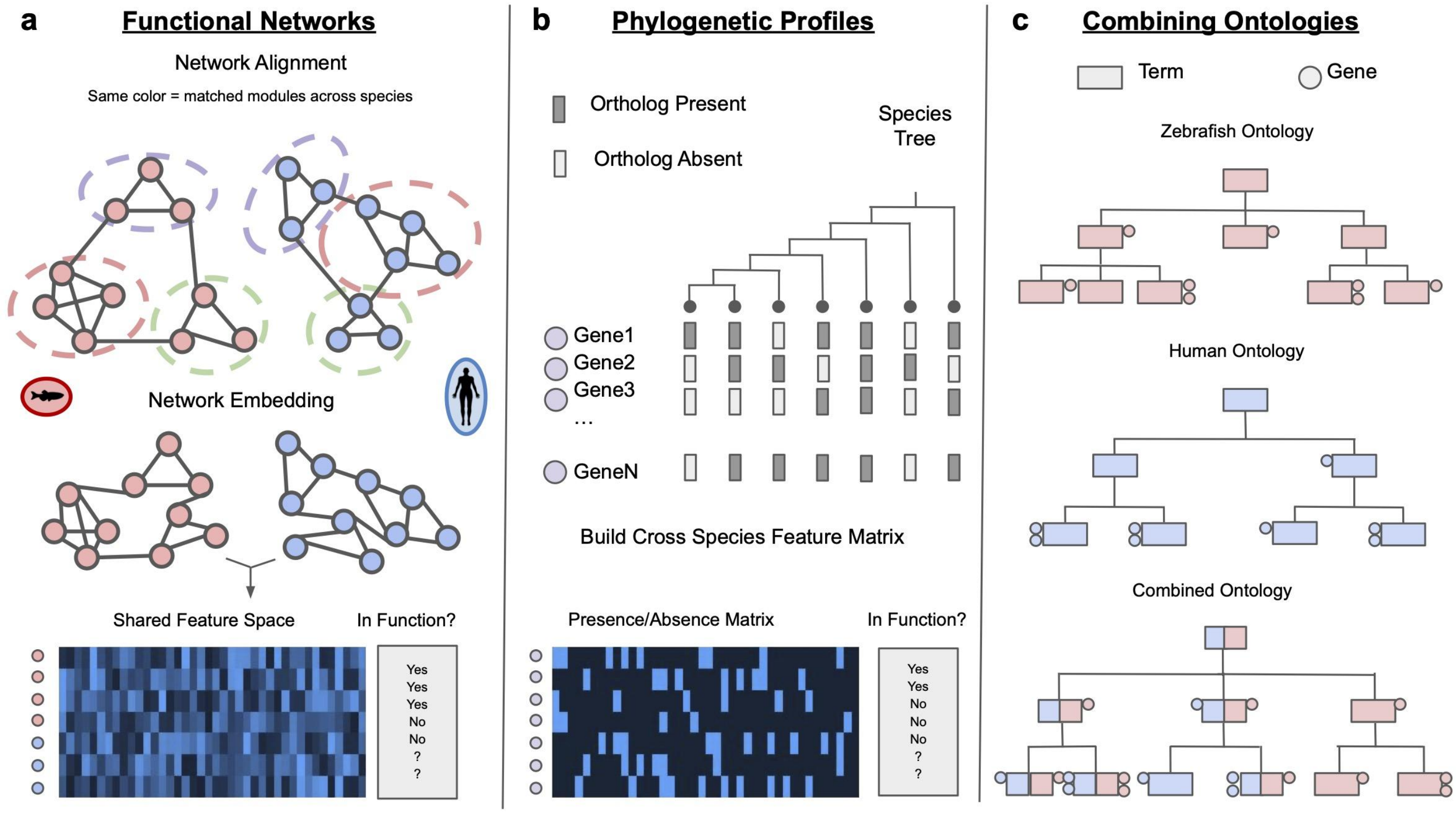

**Figure 2: How to predict disease-gene or function-gene relationships across species. (a)** Network-based methods can annotate diseases or functions for genes across species by aligning networks or by embedding networks into a shared, low-dimensional feature space where a supervised learning model is then trained to propagate annotations. **(b)** Phylogenetic profile-based methods identify co-presence or co-absence of genes throughout evolutionary history, implying closely related functions among these genes. This relationship is then used to propagate annotations across species. **(c)** Disease and function annotations can be transferred by combining ontologies across species. Genes with similar functions or disease associations across species can then be identified through the ontology structure.

## Utilizing molecular networks to predict functional and disease gene annotations across species

Molecular interaction networks, which map experimentally or computationally derived relationships between genes or their products, enable the prediction of gene functions and disease associations through the "guilt-by-association" principle, which posits that genes interacting within shared network neighborhoods are likely involved in similar biological processes[35–37]. Interaction neighborhoods within these networks capture the "functional context" of genes, which is highly complementary to sequence homology information. Further, these networks typically contain and therefore enable making inferences about nearly all genes in the genome, including non-homologous and un(-der)studied genes. Therefore, a number of network-based computational approaches have been developed to transfer and predict function, phenotype, and disease annotations across species, especially using machine learning (ML) approaches[38–41] (Figure 2a).

For instance, the Functional Knowledge Transfer (FKT) method[42] first finds "functionally-similar homologous gene pairs" that are in the same gene family and in similar network neighborhoods[40], and uses these pairs to propagate functional annotation across species. FKT successfully predicted zebrafish genes involved in mitotic exit regulation, such as cdh2, which was later experimentally validated and has then been experimentally confirmed related to cell cycle progression in zebrafish retina cells[43]. Similar to the idea in FKT, GenePlexusZoo[44] enhances cross-species predictions by simultaneously linking networks of human and five research organisms (mouse, zebrafish, fly, worm, yeast) through gene homology to generate a "functional representation" of genes from all these species, which improved prediction performance for pathway, phenotype and disease annotation tasks. Beside considering network similarity alone, NetQuilt[45] integrates multi-species networks using IsoRank[41], which aligns networks considering both sequence similarity and network similarity, and applies deep learning to predict annotations across species.

Beyond gene function prediction, network-based methods also aid in disease-gene association discovery. Katz measure[46], a well-established technique in social network link prediction[47], quantifies gene-disease proximity in heterogeneous networks combining cross-species gene orthologs and disease associations[38]. Further, "Katz features" that represent the number of paths of a certain length between a gene and a disease node, when incorporated into a ML framework, were shown to significantly enhance the performance of predicting gene-disease relationships[38]. Instead of using curated disease annotation, DiseaseQUEST[39] predicts disease-gene candidates in research organisms by combining human genome-wide association studies (GWAS)[48] with gene networks that reflect pathways underlying tissue-specific physiology[49] and disease[50,51]. Using the homologous and functionally equivalent genes of human GWAS disease genes as positive examples in a target research organism, and gene interactions in an appropriate tissue network as features, diseaseQUEST trains an ML model to prioritize novel disease-gene candidates. Strikingly, the authors demonstrated this approach by prioritizing and then experimentally validating genes related to Parkinson's disease in the nematode Caenorhabditis elegans[39].

In addition, numerous methods have been developed for predicting disease-gene associations in single species, typically in humans. These methods can be adapted to predict disease-gene associations across species[36,52–55].

## Discovering disease-related genes using phylogenetic profiles

Genes with similar functions tend to appear and disappear together during evolutionary history. Most human disease genes have ancient origins[56–59], with many traceable to eukaryotic ancestors and some dating back to the evolution of multicellularity[56]. By identifying genes that co-evolve with known disease-related genes, we therefore could propagate the disease annotations from the known genes to the co-evolved genes, thereby helping predict functions for genes that are not well characterized[60,61] (Figure 2b).

Maxwell and colleagues[62] used evolutionary profiles to examine the evolutionary distribution of human disease genes from the Online Mendelian Inheritance in Man (OMIM) database[63]. They revealed heterogeneity underlying the evolutionary origins of different classes of human disease genes. For example, genes related to inflammatory and immune diseases are of vertebrate origin, while genes associated with cardiovascular and hematological diseases originate from early metazoans.

Recently published methods[64,65] enabled systematic screens of genome-wide co-occurring functional modules, leading to functional predictions for numerous previously uncharacterized genes. For example, using phylogenetic profiling, Dey et al.[65] identified understudied candidate genes linked to the actin-nucleating WASH complex and ciliary and centrosomal defects by scanning co-occurring gene modules across 177 species from the eukaryotic tree of life. They further experimentally evaluated candidate functions of genes by identifying gene product co-localization and gene knockdown in cell models.

## Bridging disease-gene association through phenotypic similarity

Extensive gene-phenotype associations in research organisms have been discovered through hypothesis-driven experiments and large-scale genetics screens[66–68]. These links can lead to the identification of genes in a research organism whose perturbation results in phenotypes similar to those observed in human patients with a particular disease, thereby pointing to a viable model for the disease under observation.

However, comparing phenotypes across species is challenging due to evolutionary divergence and non-standard descriptions of phenotypes in different species. The continuous efforts in phenotypic ontology curation and the development of cross-species phenotype ontologies such as uPheno[69] and PhenomeNET[70] have now significantly mitigated this challenge, enabling researchers to use ontology-based semantic similarity measurements between phenotypes to explore suitable research organisms and identify new disease-gene and function-gene relationships (Figure 2c). For instance, Exomiser[71] leveraged disease-to-gene relationships discovered through semantic similarity measures among integrated ontologies to assist in disease diagnosis.

Recent studies improve on semantic similarity by creating a latent embedding space based on the phenotype ontology combined with known disease-phenotype and disease-/phenotype-gene associations[72,73]. “Node embeddings” created this way contain a low-dimensional numerical representation of each entity, such as gene, phenotype, and disease that captures that entity’s relationships. Such embeddings naturally lend themselves as inputs into ML algorithms. In this study[72], the authors trained a supervised neural network model to predict gene-disease associations based on the node embeddings of genes and diseases, which performed better than an unsupervised approach based on the similarities between gene and disease embeddings.

# How to identify functionally equivalent molecular components across species

Complementary to inferring genes in each species that are associated with a particular function or disease, another critical task is to infer molecular components that are functionally equivalent counterparts of each other across species. A number of computational approaches have been developed to identify such functionally equivalent components at the gene, pathway (gene set), and genomic level across species (Figure 3). While the concept of identifying functionally equivalent genes partially overlaps with gene-level predictions introduced in the previous section, this section expands on describing functional equivalency between higher-order functional elements.

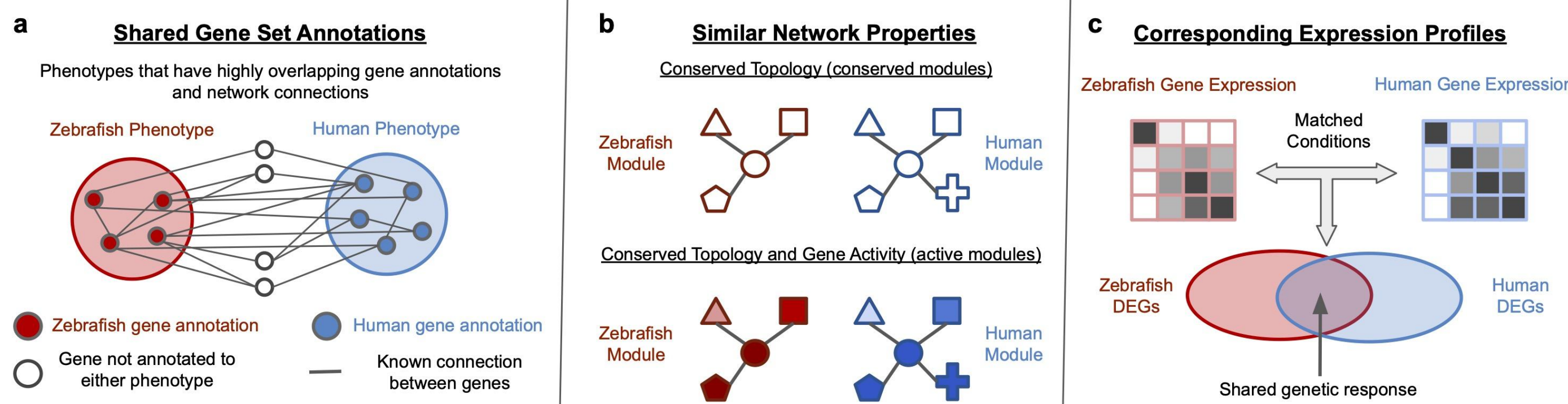

**Figure 3: How to identify functionally equivalent molecular components across species. (a)** Functionally equivalent gene pairs can be identified through genes that connect to similar sets of genes in aligned cross-species gene networks. **(b)** When prior knowledge about gene sets is known, functionally equivalent gene sets are discovered through large overlaps between gene sets across species. When prior knowledge is not available, functionally equivalent gene sets can be discovered by finding sub-networks with similar topology and expression patterns across species. **(c)** Resemblance at the organismal level, i.e., functionally equivalent profiles, can be examined by comparing genomic profiles across species.

## Identifying functionally equivalent gene pairs across species

An essential ingredient of finding functionally equivalent genes is to go beyond solely using sequence similarity and homology because, in many organisms, proteins performing similar functions (e.g., playing similar roles in the same biological pathway) may not be the most similar in sequence or share a common evolution origin[74]. Instead, equivalent molecular functions may have convergently, i.e., independently evolved. Gene Analogue Finder implemented this notion by measuring functional similarity between a pair of genes based on the overlap between gene ontology (GO) terms associated with them[75]. Han et al.[15] used a comparable approach to calculate functional similarity between orthologous human-mouse gene pairs based on the average pairwise similarity between human and mouse phenotype ontology terms annotated to those genes. This study identified several cases of functional divergence of orthologs that could be traced to changes in noncoding regulatory sequences of gene pairs with high protein sequence similarity. Nevertheless, the performance of such methods depends heavily on the completeness of gene annotations to terms in function and phenotype ontologies, with term overlap estimates becoming less meaningful for particular genes or entire species with sparse (i.e., highly incomplete) annotations[15].

Genome-scale gene interaction networks help overcome this limitation by providing a powerful alternative way to capture the “functional context” of each gene in terms of its local network neighborhood. Thus, two genes compared across species interacting with similar sets of genes in their respective molecular network neighborhoods are likely functionally equivalent (Figure 3a). Chikina et al.[40] realized this idea by first grouping the network neighbors of individual genes into meta-genes that correspond to Treefam families[76] and measuring the hypergeometric overlap between sets of meta-genes that are neighbors of a pair of genes across species in the respective species network. Intersecting meta-genes identified through this approach revealed specific biological processes underlying network similarities.

Other recent network-based approaches for finding functionally equivalent genes take advantage of the idea of node embedding. Methods such as MUNK[77], MUNDO[78], and ETNA[79] create a joint network-based representation of genes by embedding the molecular networks in a pair of species individually and then align the two embeddings based on sequence orthologs. MUNK produces different joint embeddings based on the “source” species and the “target” species, where the size of joint embedding set is as large as the number of genes in the smaller network. ETNA uses cross-training to align the two embeddings into a bidirectional compressed (low-dimensional) space. MUNK, MUNDO and ETNA use the similarity of embedding vectors of genes to estimate their functional similarity or to inform function or disease gene prediction.

Another promising approach to identify functionally equivalent genes involves comparing protein embeddings generated by protein language models (PLMs), such as ESM2[80]. These embeddings implicitly capture protein structure, molecular characteristics, as well as evolutionary relationships between proteins[81]. Protein embeddings also enable convenient comparison of sequences across diverse species using simple cosine similarity, independent of sequence similarity. This strategy has led to recent methods, including those by Hamamsy et al.[82] and Hong et al.[83], which utilize PLMs to efficiently identify deeply homologous proteins across diverse species, even in cases of limited sequence similarity. Although these methods primarily focus on identifying homologous protein domains, they provide an important foundation for identifying structurally and functionally equivalent or similar genes across species. By encoding complex DNA sequence patterns into embeddings, DNA foundation models such as Evo2[84] also offer complementary approaches for identifying functionally equivalent genes across species.

## Discovering functionally equivalent gene sets across species

Going beyond individual gene pairs, it is also of interest to find functionally conserved gene sets across species because sets of genes could represent concepts like pathways and molecular mechanisms underlying conserved phenotypes. McGary et al.[85] introduced the concept of orthologous phenotypes, i.e., “phenologs”, defined as phenotypes that are associated with orthologs. Phenologs could include phenotype counterparts that are observably very different from each other while being influenced by conserved molecular functions (e.g., significantly overlapping sets of orthologous genes). Examples include a yeast model for angiogenesis defects, a worm model for breast cancer, mouse models of autism, and a plant model for human neural crest defects[85]. Phenologs serve as a valuable tool to quantitatively identify non-obvious phenotypes in research organisms for studying human diseases, along with disease gene candidates such as genes annotated to non-obvious phenotypes and which are not orthologous to any known disease genes. MUNK[77] removes the restriction of comparing only the overrepresentation of orthologous genes. Instead, it identifies phenologs by detecting functionally equivalent genes with similar protein embeddings, uncovering more phenologs than McGary et al.

## Discovering functionally equivalent gene modules across species

Phenolog approaches rely on prior knowledge of functional and phenotypic annotations of genes, which is often highly incomplete, especially in non-traditional model species. Network-based methods have proven valuable in overcoming this limitation by helping to find conserved gene sets, usually called gene modules, while filling in knowledge gaps. This is because, in addition to capturing relationships between pairs of genes, networks also capture higher-level organization between groups of genes in the form of tight sub-networks. Consequently, similar sub-networks across species correspond to homologous functional modules or pathways (Figure 3b). CoCoCoNet[86] takes such a network-based approach to test whether a given gene set in one species is involved in similar functions as homologous gene sets in another species. If a subset of genes accurately predicts the remaining genes in the gene set using neighbor voting in both species, that is taken as an indication that the gene set corresponds to a conserved functional module.

Molecular networks can further refine comparisons of differential gene expression across species to help identify functionally conserved “active modules” that comprise tightly connected sub-networks of conserved genes that respond similarly to a given condition (e.g., disease, perturbation). However, finding such modules is difficult because active modules in different species are often not conserved across species, while conserved modules are not necessarily active[87]. So, algorithms for finding conserved and active modules need to consider activity plus conservation at the same time.

The neXus algorithm[88] was developed to meet this need. neXus identifies modules using a greedy seed-and-extend algorithm. It begins with a pair of orthologous nodes as seeds and iteratively extends both sub-networks by incorporating neighboring orthologous gene pairs. ModuleBlast[89] works similarly to neXus with the added capability of distinguishing the resulting modules based on the direction of the expression change. This separation allows ModuleBlast to evaluate whether conserved active modules display expression patterns in the same or opposite way. Both neXus and ModuleBlast limit the identification of modules to fully conserved genes. xHeinz[87] relaxes this constraint by allowing users to define the proportion of conserved nodes, offering a more flexible approach that leads to including functionally conserved but non-homologous genes in the final modules. xHeinz was further applied to understand the regulatory mechanism of interleukin-17-producing helper T (Th17) cell differentiation in humans and mice. The study revealed that key regulators of Th17 cells are conserved across species[87].

## Measuring equivalent profiles between human and research organisms

Beyond genes and gene modules, evaluating the ability of a research organism to recapitulate specific human phenotypes or disease conditions is crucial for functional studies and experimental design. However, such evaluations are subjective, with researchers often using vague terms such as “poorly” or “greatly” to indicate resemblances. Therefore, some recent methods have leveraged functional genomics data to draw quantitative conclusions about the biological resemblance between human and research organisms (Figure 3c).

Congruence Analysis for Model Organisms (CAMO)[90] quantitatively measures the congruence between human and research organisms by comparing the distribution of differentially expressed gene (DEG) profiles under matching conditions. To improve the accuracy of identifying DEGs, CAMO infers differential posterior probabilities of genes based on p-values from conventional pipelines. The concordance level of differential gene expression across species was summarized as concordance (c-scores) and discordance scores (d-scores), which served as quantitative measures of congruence across species. CAMO was applied to reconcile studies[91,92] that reached contradicting conclusions about whether mice are a suitable model for studying human inflammatory diseases. By reanalyzing and comparing inflammatory expression data within and across species, the authors concluded that burn- and infection-induced inflammation in mice resembles human inflammation[90].

Comparing phenotypes provides another avenue to discover biological resemblance in research organisms. Cross-species phenotypic ontologies can link phenotypes related to human diseases to research organism phenotypes. Leveraging such connections, the Monarch Initiative[93] provides tools, such as the Phenotype Explorer, to prioritize research organisms based on phenotypic similarities between phenotype terms from human and research organisms, aiding in the discovery of relevant research organisms and phenotypes.

## How to infer perturbed transcriptomes across species

Bulk and single-cell transcriptome profiling have emerged as preeminent technologies for nearly any organism to capture genome-wide molecular responses to a variety of developmental and physiological states as well as treatments, perturbations, and other conditions. Transcriptome profiling in research organisms has especially been valuable in studying perturbations that may be impractical or ethically infeasible in humans. Further, comparing transcriptomic profiles across species sheds light on conserved and distinct cellular states and gene responses, and makes way for context-specific knowledge transfer. However, directly comparing expression changes across species based on gene homology is challenging due to evolutionary divergence in gene expression programs. This section describes several methods that have been developed to utilize computational techniques to enable accurate comparisons of transcriptomic profiles across species (Figure 4).

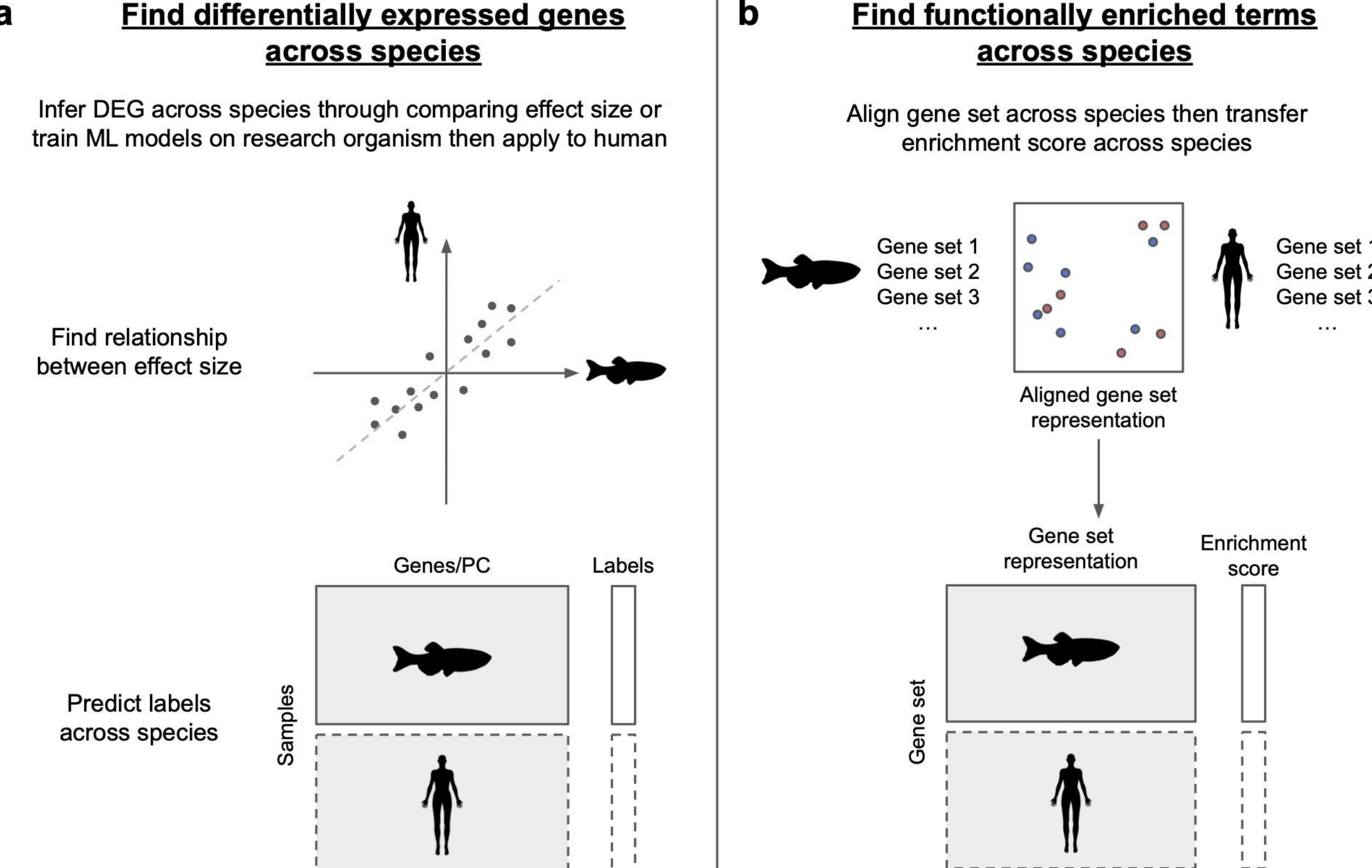

**Figure 4: How to infer perturbed transcriptomes across species.** **(a)** Some methods use linear models to relate DEG effect sizes utilizing matched samples between humans and research organisms. Other approaches develop ML models based solely on research organisms to predict conditions in humans, which is then used to identify DEGs between conditions. **(b)** To infer enriched gene sets across species, gene set representations are aligned, then models are built that can transfer enrichment scores between species.

## Determining differentially expressed genes across species

Identifying differentially expressed genes (DEGs) from transcriptomic profiles has led to insights into the impact of numerous diseases, perturbations, or experimental conditions. To systematically capture the relationship of DEGs across species, a class of methods train ML models using carefully constructed cross-species dataset pairs (CSDPs) with matching conditions, offering curated examples of how expression changes correspond to phenotypic outcomes across species, which can be further used for model training[94,95] (Figure 4a).

Found In Translation (FIT)[94] is a linear regression model that is used to understand relationships of DEGs between human and research organisms for each orthologous gene pair. The authors built models based on 170 mouse-human CSDPs across 28 conditions, which were pairs of mouse and human experiments associated with the same disease or condition. The model was then used to predict DEGs in humans under conditions analogous to those captured in novel mouse data. FIT identified more human DEGs compared to simple transfers of DEGs from research organisms to human based on orthology. For instance, using protein immunostaining, the authors confirmed their prediction that the gene ILF3 is upregulated in the colon of patients with inflammatory bowel diseases (IBD) even though the gene was not differentially expressed in either human or mouse data[94].

The FIT approach is likely to be limited to diseases or drug perturbations with sufficient training data from both human and research organisms. In contrast, Brubaker et al.[95] proposed a semi-supervised method to predict DEGs across species that only requires phenotype information in research organisms. Initial supervised models were built to predict mouse phenotypes using mouse expression data, and then these models were iteratively augmented with high-confidence human samples to predict the phenotypes of the remaining human expression data. Finally, differential gene expression analyses were performed on human samples with distinct predicted phenotypes.

Transcomp-R[96] was developed to predict DEGs across species when phenotype labels are available for only one species, which applied to identify mouse genes predictive of infliximab responses in IBD patients, despite the absence of corresponding phenotype labels in mice. By applying principal component analysis (PCA) to murine expression data and projecting human expression data into the murine PC space, TransComp-R identified murine PCs and genes associated with infliximab response. This approach highlighted integrin signaling activation in resistant patients, with single-cell sequencing revealing ITGA1 overexpression in immune cells. The role of ITGA1 in resistance was further validated through anti-ITGA1 treatment on patient immune cells. Beyond infliximab response in IBD, TransComp-R has been used to identify other translatable signatures, such as MK2 kinase inhibition in IBD[97] and common disease signatures among mouse and human in Alzheimer's disease[98]. Similar projection techniques have also been applied to cross-species pathway comparisons[99].

Finally, other recent work[100,101] have treated biological conditions—such as tissue type or disease state—as a "style" and applied style transfer techniques to predict transcriptomic changes within a species. AutoTransOP[102] extends this concept to cross-species comparisons by treating species as a style. Using an autoencoder structure, AutoTransOP compresses gene expression profiles into a latent space, maximizing mutual information and minimizing the distance between same-condition embeddings to create a meaningful representation. When a new sample from one species is introduced, the model can predict its corresponding expression profile in another species. Interestingly, AutoTransOP also modified TransComp-R for cross-species profile prediction by using PCA as a latent variable and decoding it to the translated profile. In translating lung fibrosis profiles between humans and mice, TransComp-R achieves performance comparable to AutoTransOP when orthologous genes are sufficiently available, suggesting that linear transformations may be adequate in such contexts. However, when orthologs are limited, AutoTransOP excels.

Though promising, it is important to note that genetic differences across species are significantly larger and more complex than the condition changes observed within a single species. As a result, applying style transfer techniques directly to cross-species comparisons may face additional challenges and limitations that need to be carefully considered and addressed.

## Determining functionally enriched terms across species

Gene set analyses (GSA) have been widely used to identify predefined gene sets (e.g. Gene Ontology biological processes) that are significantly enriched in a gene list of interest[103]. GSA provides a powerful approach to gaining insights into enriched functional terms associated with diseases, phenotypes, or perturbations. However, the discrepancy of genes across species makes the definition of functionally equivalent gene set across species difficult, so directly transferring enriched terms across species is impractical (Figure 4b).

XGSEA[104] tackles the challenge of transferring enriched terms across species by using affine mapping, which projects genes from humans and research organisms into the same space. In this joint space, gene sets sharing more homologous genes will be closer to each other in the transformed space. With the transformed gene set representations, logistic regression models were built to characterize the relationship between the representation and enrichment metrics from the research organism, including p-values and enrichment scores of gene sets. Finally, the trained models are used to predict enrichment metrics for gene sets from humans. XGSEA successfully estimates enrichment metrics in human gene sets based on metrics of homologous gene sets in research organisms. For instance, XGSEA was able to train computational models based on zebrafish data and predict enriched pathways associated with melanomas in human patients[104].

# How to map equivalent cell types and cell states across species

The advent of single-cell and single-nucleus sequencing techniques has opened up new avenues in biological research. This emerging frontier focuses on identifying and comparing cell types and states across different species to better understand cellular diversity and cellular innovations that arose through cell type evolution[105,106]. By transferring insights from extensively studied organisms to less-explored ones, these cross-species comparisons offer a powerful insight for cell type identification and discovery, thereby advancing our understanding of cellular biology across species.

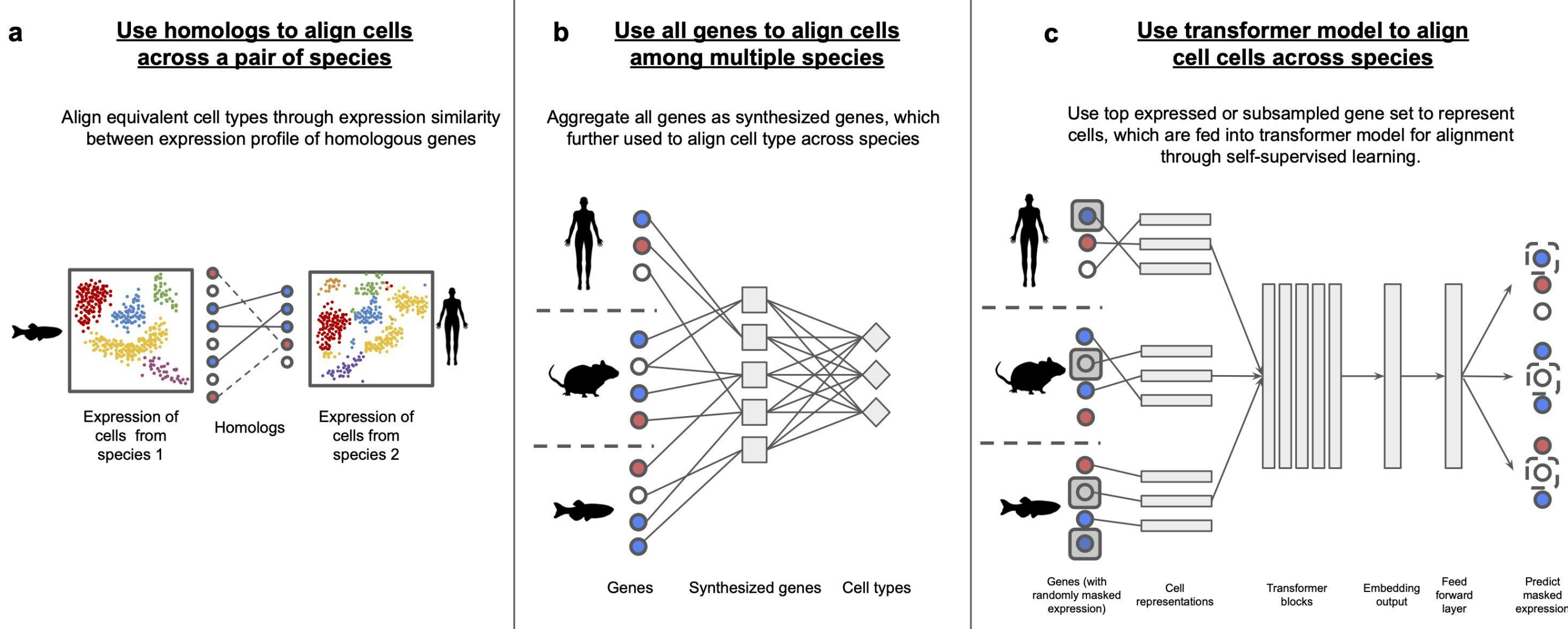

**Figure 5: How to map equivalent cell types and cell states across species. (a)** The majority of methods align cells with similar expression patterns of homologous genes across a pair of species. **(b)** SATURN stands out as the only non-

transformer-based method capable of aligning single-cell datasets from multiple species simultaneously. It first aggregates genes into synthesized macrogenes using protein language model. These synthesized macrogenes are then utilized to align cells across multiple species. **(c)** Newer methods are employing transformer architectures for self-supervised alignment of cells across species. Genes are first subsampled – either selecting top-expressed genes or random subsets – and then embedded to represent cells. These embeddings pass through transformer blocks followed by a feed-forward layer to predict the expression level of masked genes from the input or whether the masked genes are expressed in the cells.

Cell mapping problems may be simply conceptualized as mapping transcriptionally similar cells across species (i.e., based on homologous genes having similar expression levels). However, cells of a particular type within a species are more alike than those of corresponding cell types across different species[106,107]. Therefore, effectively mapping cells across species requires correcting for species-specific expression differences so that cells of homologous types are mixed in irrespective of species origin, while cells of different types remain distinct from each other. Additionally, methods must account for experiment-induced batch effects, which arise from technical variations rather than true biological signals. Some tools have been developed to reconcile heterogeneous single-cell RNA-seq (scRNA-seq) data from multiple species[108,109]. These methods typically seek to project cells from multiple species into a unified space, facilitating cross-species comparison. As a result, the cell mapping problem can be viewed as a domain adaptation problem, i.e., aligning cells from different species in a unified space (Figure 5).

## Aligning cells across a pair of species based on homologs

Shafer[106] summarized various methods for cross-species scRNA-seq integration. Originally designed to mitigate the batch effects that were introduced by experimental variation, these methods may not sufficiently correct for species differences, which are more pronounced than batch effects[107]. When comparing human cells to those of anciently polyploid species like the teleost fishes (e.g. zebrafish), the abundance of gene duplicates results in fewer one-to-one gene matches, leading to a reduced signal for alignment. The reliance on orthology stems from the orthology conjecture, which posits that orthologs (i.e., homologous genes in different species that evolved from a common ancestral gene by speciation) are generally the most functionally similar genes across species, while gene duplicates (paralogs) tend to functionally diverge more at equivalent sequence distances. However, orthology does not necessarily imply similar functions across species, and paralogs might exhibit more functional similarity than orthologs. For instance, when an ortholog acquires a loss-of-function mutation, its function might be compensated by the upregulation of a paralog, resulting in the paralog having a more similar function to the ortholog in the other species[110,111]. Therefore, incorporating one-to-many, many-to-one and many-to-many homology into the analyses could improve cross-species alignment performance by accounting for functional similarity between paralogs (Figure 5a).

In a benchmark study conducted by Song et al.[112], SAMap[113] was identified as the only method that is capable of mapping divergent cell types. To account for the effect that gene expression patterns may be divergent in homologous cell types, SAMap uses one-to-one, one-to-many, many-to-one and many-to-many homology information and incorporates neighboring cells within species into the calculation of similarity between cells across species. Consequently, cells with differing expression profiles remain closely associated if they are among the nearest cross-species neighbors to each other. Based on this analysis, SAMap identified a general alignment of gene expression patterns and developmental relationships during embryogenesis in frogs and zebrafish. Interestingly, they also detected a group of secretory cell types that have similar expression patterns while having different developmental origins in the two different species including cells arising from different germ layers[113].

## Aligning cells among multiple species using all genes

Instead of using precalculated gene homology, the new framework SATURN[114] integrates cell atlases from divergent species by harnessing the power of protein language models (PLMs) to relate genes across species to each other and project all cells from multiple species into shared cell embedding space (Figure 5b). Using a neural network, SATURN first projects scRNA-seq datasets to a joint space composed of "macrogenes" representing groups of genes across species that have similar protein embeddings inferred from the PLM. The neural network weights are used to define how a gene contributes to a macrogene. For each macrogene, SATURN learns the cross-species cell embedding as a non-linear combination of macrogenes, guided by an unsupervised objective function that simultaneously maximizes the distance between distinct cells within the same species and minimizes the distance between similar cells across species. This approach enables SATURN to integrate data from multiple species (and not just perform pairwise comparisons), including divergent species where gene homology relationships are hard to unravel. The output of SATURN also enables easy cross-species comparison, such as finding differentially-expressed macrogenes across species, which can then be traced back to actual genes based on the weight of connections to the macrogenes in the neural network. SATURN's application to a multi-species dataset of frog and zebrafish embryogenesis shows success in aligning evolutionarily related cell types and revealing differentially expressed genes in macrophage/myeloid progenitors and ionocytes across these two distantly related species[114].

## Advancing cross-species cell type alignment with transformer models

A new wave of approaches for aligning cell types across species is based on transformer models, which take large-scale single-cell data from multiple species and tissues as input and are trained in a self-supervised manner (Figure 5c). A key challenge in these transformer models is encoding cells with different numbers of genes across species into a fixed-length token sequence due to token limitations. Another issue is determining the gene ordering within a cell. Following solutions from single-species single-cell transformer models[115,116], cross-species models like Precious3GPT[117] and GeneCompass[118] encode cells as sequences of gene embeddings ordered by expression level. They also add special species tokens to differentiate cells from different species. However, as the typical maximum token length is limited, lower-expressed genes, which may have important functions, are often ignored. The Universal Cell Embeddings (UCE)[119] model addresses this limitation by subsampling a fixed number of genes based on their expression level, allowing it to capture expression patterns from genes with low-expression. UCE also applies an alternative gene ordering strategy, sorting genes by genomic location to capture gene regulatory and gene order (synteny) relationships that are often evolutionarily conserved, aiding cross-species integration.

Precious3GPT and GeneCompass models include all human genes but only homologs from other species. This reliance on homology limits these models to a few closely related species. On the other hand, instead of learning gene embeddings from scratch, UCE[119] uses protein sequence embeddings as input, similar to strategies applied in SATURN[114]. While this approach restricts the model to protein-coding genes, it bypasses homology constraints and enables easier and broader cross-species comparisons. Some models integrate prior knowledge with the expression data to improve performance. For example, GeneCompass[118] incorporates information on gene regulatory networks, gene promoters, gene family annotations, and gene co-expression relationships. Precious3GPT[117] enhances predictions using knowledge graphs and text description of genes from National Center for Biotechnology Information (NCBI).

These models provide a way to leverage insights from model species to better understand non-model species through predictions without requiring prior knowledge (i.e., zero-shot predictions) or relying on homology. With the number of cross-species single-cell models steadily growing, it is critical to perform comprehensive benchmarking to evaluate their ability to learn complex expression patterns across species, assess integration quality, and ensure accurate cross-species cell-type predictions.

## Future perspective

The explosion of large-scale human biological data and accompanying analysis methods, while seen as a prospect of a golden age for human genetics, has also raised the question of whether it will diminish the significance of research organisms[7,120]. From the perspective of this review, the answer is certainly "No". In fact, the ocean of new data will lead to an increasing number of candidate disease-related genes, dysregulated pathways, drugs, etc., that will require validation and functional characterization, which can be carried out comprehensively only in the in vivo settings of research organisms[7]. Therefore, the goal of increasingly better computational strategies to effectively gain biological and biomedical insights from research organisms will continue to remain significant and grow further. The landscape of current methods, summarized in this review, demonstrates great promise towards this goal. Instead of solely relying on sequence similarity and gene homology, these approaches utilize sophisticated inferential techniques and diverse datasets, paving a promising path for effectively transferring knowledge between human and research organisms. The next frontier in unlocking the full potential of research organisms and the ever-growing datasets involves addressing some key challenges that still remain.

### Capturing the specific facets of complex diseases

Complex human diseases exhibit staggering genetic and phenotypic heterogeneity. Despite this complexity, much of the current focus is on identifying a single research organism that can fully recapitulate a disease condition. Consequently, a major need in the field is to develop methods that can identify optimal phenotypes, conditions, and genes for studying the many facets of complex diseases.

Several approaches have been proposed to dissect diseases into molecular components. Given that disrupted biological processes are often shared among diseases[121–123], even those that are seemingly unrelated[124], it is promising to dissect diseases into dysregulated processes[125] or modules[126,127]. Additionally, phenotype ontologies, such as Human Phenotype Ontology (HPO)[128], are widely used tools for phenotype-driven disease analyses, enabling the breakdown of diseases into related phenotypes[129,130].

By leveraging dissected processes, modules, or phenotypes, we can identify combinations of research organisms that maximally capture the multifaceted nature of complex diseases. The methods discussed in this review, particularly in the section "How to identify functionally equivalent molecular components across species", shed light on approaches for relating these dissected components to suitable research organisms. However, there remains a critical need for future approaches that can utilize functionally equivalent molecular components of complex diseases to guide the strategic selection of the most appropriate research organisms for specific disease aspects.

### The concept of agnology: sometimes we just don't know

Homology is widely used as the bridge to find functional similarities across species. However, the definition of homology, i.e., shared evolutionary origin of a gene or phenotype, in itself does not include function nor does homology guarantee functional similarity. Many factors can lead to functional divergence between homologous genes. Examples include changes in non-coding regulatory sequences[15], reciprocal gene loss[131], and developmental system drift[132].

Moreover, solely focusing on homology restrains the power of predictive models to explore comprehensive functional relationships across divergent species. For instance, research organisms like the nematode C. elegans may not share a substantial number of clear and direct gene orthologs with humans, resulting in a scarcity of genes available for model training and testing. This limited orthologous gene repertoire can hinder the ability of computational models to learn comprehensive functional relationships across species. Furthermore, orthologs alone can only explain a fraction of observed biological variations[106]. Therefore, it is crucial to also incorporate non-homologous genes to construct more comprehensive models. Relying solely on homologous genes is also limiting from the perspective of network biology, where species evolve through gain or loss of interactions, indicating that non-homologous genes can play roles in similar functional and regulatory programs[13][3].

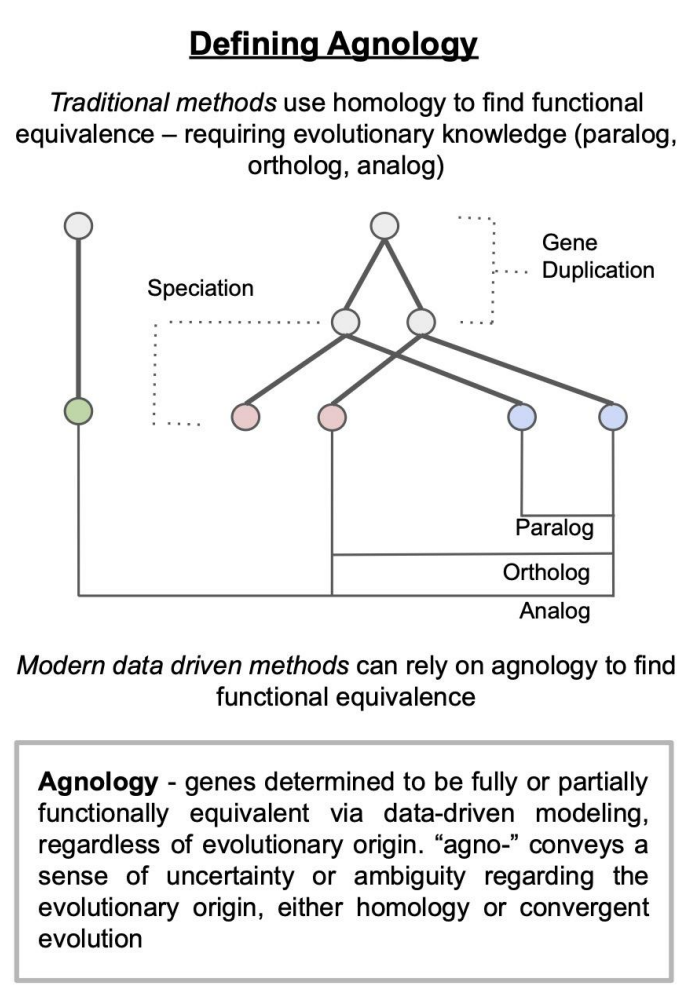

**Figure 6: Definition of agnology.** Traditional methods rely on homology to determine functional equivalence. In contrast, modern data-driven approaches introduce agnology, identifying fully or partially functionally equivalent genes regardless of evolutionary origin, allowing for ambiguity in homology or convergent evolution/analogy.

Some methods discussed in this review, like FIT[94] for cross-species profile prediction or CoCoCoNet[86] for network module comparison, focus mainly on homologous genes. However, there is a growing shift toward broader comparisons beyond homology. Some approaches, like MUNK[77] and GenePlexusZoo[44], use homology as anchors to expand comparisons to non-homologous and species-specific genes. Others, like SATURN[114] and UCE[119], directly use protein embedding similarity to create a more expansive map between genes across species without being restricted to a priori inferred gene homologies. To formally describe computationally discovered fully or partially functionally equivalent genes that may or may not be homologous, i.e., when "we just do not know", we introduce the term "agnolog" (Figure 6). Agnologs are biological entities, processes, or responses—such as genes, gene sets, or biological systems—that are functionally equivalent across species regardless of their evolutionary origin. The prefix "agno-" means "unknown" or "not known", reflecting a data-driven observation, independent of evolutionary assumptions.

In this context, agnology bridges the gap between homology and analogy. Analogs are functionally similar entities that evolved independently, while homologs share a common ancestry, though some may have diverged functionally. Agnologs encompass both analogs and homologs with similar functions. Identifying agnologs allows for further evolutionary analysis to determine whether they are in fact analogs or functionally equivalent homologs.

Gene rescue experiments provide a potential way to determine agnologs, where a gene is functionally disabled in one species and replaced by a gene from another species to assess whether the original phenotype can be restored. These assays, commonly performed in research organisms like

Drosophila[134] and yeast[135], have traditionally focused on orthologous genes to reduce the search space. Methods discussed in this review can help to broaden candidates to non-orthologous genes. For example, MUNK predicted human gene sets associated with muscular dystrophy to be functionally related to mouse gene sets linked to abnormal muscle fiber morphology, despite minimal shared homology. Specifically, MUNK identified 21 functionally similar protein pairs between these phenotypes, with only one being homologous[77]. Another representative example is GenePlexusZoo, which predicted genes associated with Bardet-Biedl Syndrome 1 (BBS) across five research organisms. Notably, among the predicted genes annotated to top BBS-related biological processes in these organisms, none of the genes turned out to be orthologous to the human BBS genes[44].

Similar to homology being defined at different levels[136], agnologs can be defined and analyzed at multiple biological levels, including genes, gene sets, and cell types. At the gene level, agnologous pairs are genes with similar functions across species, typically identified by first anchoring a subset of genes through homology, followed by exploration of functional similarities to all genes beyond homologous relationships. These agnologous genes are often recognized using embedding representations, enabling systematic comparisons. At the gene-set level, agnologous gene sets are groups of genes responsible for similar functions or phenotypes, identified through overrepresentation analyses of agnologous genes or through predictive approaches, relying less on homology.

Identifying agnologous cell types across species represents a challenging yet exciting frontier in cell type evolution studies. Methods leveraging homologous genes include SAMap[113], Precious3GPT[117], and GeneCompass[118], whereas SATURN[114] and UCE[119] utilize agnologous genes identified through protein language models. For instance, SATURN identified early-stage zebrafish macrophages and frog myeloid progenitors as agnologous cell types, as both share the capability of differentiating into macrophages regulated by conserved mechanisms.

These concepts – agnologous genes, gene sets, and cell types – are deeply interconnected. Agnologous genes serve as a backbone to guide predictions of agnologous gene sets, as demonstrated by methods such as MUNK and GenePlexusZoo. Similarly, embedding representations derived from agnologous genes facilitate the identification of agnologous cell types across diverse species, such as in SATURN or UCE.

## Networks in more species and more contexts

Molecular networks are one of the most widely applied data types in cross-species knowledge transfer. Regardless of their representation (e.g., edgelists, adjacency matrices, or node embeddings), networks capture interactions among neighboring genes and provide mechanistic representations of genes to computational models. However, network-based methods also have limitations.

Most networks generated using experimental approaches are for humans or popular research organisms like mice. There is an urgent need to expand the repertoire of networks beyond these species to include non-traditional research organisms. Current methods, such as STRING[137] and those described in recent literature[138], use orthology information to infer "interologs", i.e., conserved interactions between pairs of proteins that have interacting homologs in another organism[139,140]. For example, STRING utilizes high-confidence networks in humans and data-rich research organisms to derive protein-protein interaction (PPI) networks for over 1,000 species. However, the assumption that paired orthologous genes have conserved interactions across species may not always hold true due to interaction rewiring over the course of evolution. Additionally, limiting the scope to orthologous genes precludes inferring interactions involving taxon-specific genes. Other previously introduced methods like FKT/IMP[42,141] use homologous genes with similar network neighborhoods to transfer knowledge across species, but these methods require experimental functional genomics data as prior to generating networks, limiting their application to popular research organisms.

Another important future step in generating species-specific networks is the development of context-specific networks, particularly those with tissue or cell-type specificity. Context specificity plays a crucial role in biomedicine since disease-gene associations frequently arise from disrupted interactions among tissue-specific and cell lineage-specific processes under particular environmental conditions[51,142]. Unfortunately, the nuanced interactions that vary across tissues may not be fully captured by experimentally generated large-scale networks such as PPI. Coexpression networks have the potential to capture tissue specificity more effectively[143], but they tend to be noisy[144]. To obtain robust signals, current studies often extract only a small fraction of information for downstream analysis, such as the top 0.5% of co-expressed gene pairs[145]. This approach, however, results in significant loss of information. Limited efforts have been made to build tissue-specific networks by integrating multi-modal functional genomic data[39,51] or to contextualize network representations by incorporating tissue-specific expression[146–148].

Recent advances in sequence-based deep learning models could help expand the repertoire of species-specific and context-specific networks. For example, AlphaFold-Multimer[149] can predict protein interactions based on sequences. ExPecto[150] and ExPectoSC[151] can predict tissue-/cell-specific regulatory landscapes based on DNA sequences alone. Thanks to the advancement of genome projects for non-model species, such as the Vertebrate Genomes Project[152], genome sequence resources are much richer than other genomics types such as transcriptomes or epigenomes. These sequence-based models and transfer learning techniques pave a promising way to expand networks beyond popular research organisms. However, we need to integrate phylogenetic information into transfer learning to consider the taxonomic-specific logic of gene regulation or protein interactions.

## Automated construction of ontologies and knowledge graphs

In this review, we have described the power of ontologies as a framework for transferring knowledge across species. Furthermore, combining various ontologies and annotations into knowledge graphs can provide new insights for translational biomedicine. The Monarch Knowledge Graph exemplifies this approach, integrating knowledge from 33 biomedical resources, including information from all major research organism databases[93]. Leveraging such knowledge graphs, we can employ advanced techniques like graph deep learning[153] to utilize information from research organisms and assist biomedical inquiries such as rare disease variant prioritization[154]. Integrating knowledge graphs with large language models (LLMs) can also help reduce “hallucinations” in AI-powered question-answering systems within the biomedical field[155].

Despite the power of ontologies and knowledge graphs in cross-species knowledge transfer, the generation of ontologies requires laborious curation by specialists. Manual curation may also introduce biases towards popular research areas. For instance, since zebrafish is a widely used research organism for developmental and embryogenic studies, GO terms annotated to zebrafish might be skewed towards early life stages. Such differences in terms of annotation frequency could in turn bias downstream genomic analyses[156], especially when terms are then transferred to other, less-studied species.

The capability of current LLMs to automatically extract entities and relationships from the literature offers an efficient alternative for generating ontologies and knowledge graphs. For example, SPIRE[157] can generate ontologies by processing text input and a user-provided schema that describes the desired structure of the ontology. While more effort is needed to maintain the quality and consistency of automatically generated ontologies, this approach holds great potential for significantly enriching ontologies across a wide spectrum of organisms, from well-studied to understudied research organisms and even non-model organisms.

## Better benchmarking studies for cross-species scRNAseq analyses

Mapping single-cell and single-nuclei transcriptomics (scRNA-seq/snRNA-seq) data across different species provides valuable insights into cell type evolution and facilitates cell type and cell state annotation[105,106]. However, most existing cross-species cell mapping algorithms are limited to closely related species and do not account for one-to-many, many-to-one, and many-to-many homology relationships[106], while considering all homologous genes is crucial when comparing species involving gene duplication derived from common ancestors. For example, due to a whole genome duplication in the ancestor of teleost fishes, considering all homologous gene pairs is essential for cross-species integration of cell types between human and biomedical fish models such as the teleost zebrafish[158]. To overcome such limitations, algorithms such as SAMap[113] and SATURN[114] have been developed. While recent benchmarking studies highlight properties of these methods that are important for effective cross-species integration of scRNA-seq data[112], most of these methods were originally designed for batch corrections. Furthermore, since SAMap applied additional restrictions on within-species manifolds, SAMap was not systematically compared to other methods in current benchmarking studies. Transformer-based alignment models further expand the landscape of cross-species scRNA-seq alignment methods, differing in their strategies for encoding cell embeddings, model architectures, species divergence, and the proportions of species included in the training data. However, there is currently no comprehensive benchmarking study that systematically evaluates these models. Thus, it is essential to conduct more comprehensive benchmarking of these methods to evaluate their performance on different types of scRNA-seq datasets and species with varying degrees of divergence and data curation. Additionally, the development of new methods for downstream analyses is crucial to extract meaningful biological insights from the growing mountain of single-cell/nuclei data.

## Extracting and curating knowledge from non-traditional research organisms

The majority of the organisms noted in this review are classic research organisms such as mouse, zebrafish, and nematode, with non-traditional and emerging research organisms often being overlooked in “mainstream” biomedical studies. However, non-traditional research organisms are significant for translational research due to their unique biological characteristics[159]. For instance, the axolotl (Ambystoma mexicanum), a neotenic salamander, shows remarkable regenerative abilities, making it a crucial research organism for regenerative medicine and developmental biology[160]. The spotted gar (Lepisosteus oculatus), a ray-finned fish with a uniquely slow evolutionary rate, has proven valuable in facilitating genomic comparison between teleost biomedical research organisms and the human genome[33]. The tunicate, Ciona intestinalis, is a close chordate relative of vertebrates. Its unique evolutionary position, simple body plan, and ease of manipulation make it highly suitable for studying chordate embryonic development and morphogenesis[161]. However, challenges such as polyploidy, large genome sizes, genomic rearrangements, and taxon-specific cell types can make analyzing the genomic data from non-traditional species more difficult than classic research organisms. It is thus necessary to explore additional methods that can effectively transfer knowledge and data from and to non-traditional research organisms.

## Conclusions

In this review, we have discussed the current state of computational methods for cross-species knowledge transfer. We emphasize that these methods surpass simple comparisons of molecular profiles across species and highlight the utilization of orthogonal information sources such as phenotypic ontologies and molecular networks to facilitate cross-species knowledge transfer.

Our review provides resources and insights into the advancements and challenges of methods for cross-species knowledge transfer among various research organisms and human. The resources summarized in this review will facilitate biomedical studies using research organisms, including traditional and non-traditional research organisms, by leveraging knowledge from human or extensively studied model systems.

There are still critical unmet needs and plenty of room for further improvements and refinements of existing approaches. More advanced computational approaches are needed to identify a range of research organisms for studying different aspects of diseases. Instead of relying solely on homology as a bridge, expanding analyses to agnologs can enhance the effectiveness of cross-species knowledge transfer. To gain a more comprehensive understanding of diseases, we need networks from more species, and also need to incorporate context specificity, especially cell- and tissue-specificity, into networks, despite the inherent difficulties involved. To fully unlock the power of ontology-based knowledge transfer, we need methods to automatically generate high-quality and robust ontologies as well as knowledge graphs. As a rapidly growing field, it is essential to establish more benchmarks for state-of-the-art cross-species cell mapping between humans and evolutionary divergent research organisms. Currently, most methods are focused on the few well-studied research organisms, but greater attention and analytical methods should be employed to extract biomedical insights from non-traditional, emerging research organisms.

We propose that with the application of comprehensive computational approaches, the field will gain more exciting insights from big data of research organisms, ultimately enhancing our understanding of human biology and diseases.

# Acknowledgements

This work is supported by NIH R35 GM128765 and Simons Foundation 1017799 (to A.K.). (Zebra)fish-human research transfer in the Braasch Lab has been supported by NIH R01OD011116. We thank all members in Krishnan and Braasch Labs for helpful discussion and feedback on the manuscript.

## Author contributions

H.Y. drafted the manuscript. H.Y., C.A.M, K.J., I.B., and A.K. edited the manuscript.

# Ethics declaration

# Competing interests

The authors declare no competing interests.

# References

1. Barré-Sinoussi, F. & Montagutelli, X. Animal Models are Essential to Biological Research: Issues and Perspectives. Future Sci. OA **1**, (2015).

2. Baldridge, D. et al. Model organisms contribute to diagnosis and discovery in the undiagnosed diseases network: current state and a future vision. Orphanet Journal of Rare Diseases **16**, 206 (2021).

3. Baumans, V. Use of animals in experimental research: an ethical dilemma? Gene Ther **11**, S64–S66 (2004).

4. Festing, S. & Wilkinson, R. The ethics of animal research. EMBO Reports **8**, 526–530 (2007).

5. Demers, G. et al. Harmonization of Animal Care and Use Guidance. Science **312**, 700–701 (2006).

6. Wangler, M. F. et al. Model Organisms Facilitate Rare Disease Diagnosis and Therapeutic Research. Genetics **207**, 9–27 (2017).

7. Aitman, T. J. et al. The future of model organisms in human disease research. Nat Rev Genet **12**, 575–582 (2011).

8. McClellan, J. & King, M.-C. Genetic Heterogeneity in Human Disease. Cell **141**, 210–217 (2010).

9. Tadano, R. et al. Molecular characterization reveals genetic uniformity in experimental chicken resources. Experimental Animals **59**, 511–514 (2010).

10. Franěk, R. et al. Isogenic lines in fish – a critical review. Reviews in Aquaculture **12**, 1412–1434 (2019).

11. Cui, C., Zhou, Y. & Cui, Q. Defining the functional divergence of orthologous genes between human and mouse in the context of miRNA regulation. Briefings in Bioinformatics **22**, (2021).

12. Nehrt, N. L., Clark, W. T., Radivojac, P. & Hahn, M. W. Testing the Ortholog Conjecture with Comparative Functional Genomic Data from Mammals. PLoS Comput Biol **7**, e1002073 (2011).

13. Stamboulian, M., Guerrero, R. F., Hahn, M. W. & Radivojac, P. The ortholog conjecture revisited: the value of orthologs and paralogs in function prediction. Bioinformatics **36**, i219–i226 (2020).

14. Nadimpalli, S., Persikov, A. V. & Singh, M. Pervasive Variation of Transcription Factor Orthologs Contributes to Regulatory Network Evolution. PLoS Genet **11**, e1005011 (2015).

15. Han, S. K., Kim, D., Lee, H., Kim, I. & Kim, S. Divergence of Noncoding Regulatory Elements Explains Gene–Phenotype Differences between Human and Mouse Orthologous Genes. Molecular Biology and Evolution **35**, 1653–1667 (2018).

16. Arendt, D. et al. The origin and evolution of cell types. Nat Rev Genet **17**, 744–757 (2016).

17. Büscher, T., Ganai, N., Gompper, G. & Elgeti, J. Tissue evolution: mechanical interplay of adhesion, pressure, and heterogeneity. New J. Phys. **22**, 033048 (2020).

18. Ha, D. et al. Evolutionary rewiring of regulatory networks contributes to phenotypic differences between human and mouse orthologous genes. Nucleic Acids Research **50**, 1849–1863 (2022).

19. Rhrissorrakrai, K. et al. Understanding the limits of animal models as predictors of human biology: lessons learned from the sbv IMPROVER Species Translation Challenge. Bioinformatics **31**, 471–483 (2014).

20. Zhou, N. et al. The CAFA challenge reports improved protein function prediction and new functional annotations for hundreds of genes through experimental screens. Genome Biol **20**, (2019).

21. Jiang, Y. et al. An expanded evaluation of protein function prediction methods shows an improvement in accuracy. Genome Biol **17**, (2016).

22. Radivojac, P. et al. A large-scale evaluation of computational protein function prediction. Nat Methods **10**, 221–227 (2013).

23. Friedberg, I. & Radivojac, P. Community-Wide Evaluation of Computational Function Prediction. Preprint at https://doi.org/10.48550/arxiv.1601.01048 (2016).

24. Brubaker, D. K. & Lauffenburger, D. A. Translating preclinical models to humans. Science **367**, 742–743 (2020).

25. McWhite, C. D., Liebeskind, B. J. & Marcotte, E. M. Applications of comparative evolution to human disease genetics. Current Opinion in Genetics & Development **35**, 16–24 (2015).

26. Kowald, A. et al. Transfer learning of clinical outcomes from preclinical molecular data, principles and perspectives. Briefings in Bioinformatics **23**, (2022).

27. Dougherty, B. V. & Papin, J. A. Systems biology approaches help to facilitate interpretation of cross-species comparisons. Current Opinion in Toxicology **23-24**, 74–79 (2020).

28. Feuermann, M. et al. A compendium of human gene functions derived from evolutionary modelling. Nature **640**, 146–154 (2025).

29. Wu, R. S. et al. A Rapid Method for Directed Gene Knockout for Screening in G0 Zebrafish. Developmental Cell **46**, 112–125.e4 (2018).

30. D'Agostino, Y. et al. Loss of circadian rhythmicity in bdnf knockout zebrafish larvae. iScience **25**, 104054 (2022).

31. Shearin, A. L. & Ostrander, E. A. Leading the way: canine models of genomics and disease. Disease Models & Mechanisms **3**, 27–34 (2010).

32. Castoe, T. A. et al. The Burmese python genome reveals the molecular basis for extreme adaptation in snakes. Proc. Natl. Acad. Sci. U.S.A. **110**, 20645–20650 (2013).

33. Braasch, I. et al. The spotted gar genome illuminates vertebrate evolution and facilitates human-teleost comparisons. Nat Genet **48**, 427–437 (2016).

34. Grohme, M. A. et al. The genome of Schmidtea mediterranea and the evolution of core cellular mechanisms. Nature **554**, 56–61 (2018).

35. Warde-Farley, D. et al. The GeneMANIA prediction server: biological network integration for gene prioritization and predicting gene function. Nucleic Acids Research **38**, W214–W220 (2010).

36.

Liu, R., Mancuso, C. A., Yannakopoulos, A., Johnson, K. A. & Krishnan, A. Supervised learning is an accurate method for network-based gene classification. Bioinformatics **36**, 3457–3465 (2020).

37. Wang, X., Gulbahce, N. & Yu, H. Network-based methods for human disease gene prediction. Briefings in Functional Genomics **10**, 280–293 (2011).

38. Singh-Blom, U. M. et al. Prediction and Validation of Gene-Disease Associations Using Methods Inspired by Social Network Analyses. PLoS ONE **8**, e58977 (2013).

39. Yao, V. et al. An integrative tissue-network approach to identify and test human disease genes. Nat Biotechnol **36**, 1091–1099 (2018).

40. Chikina, M. D. & Troyanskaya, O. G. Accurate Quantification of Functional Analogy among Close Homologs. PLoS Comput Biol **7**, e1001074 (2011).

41. Singh, R., Xu, J. & Berger, B. Global alignment of multiple protein interaction networks with application to functional orthology detection. Proc. Natl. Acad. Sci. U.S.A. **105**, 12763–12768 (2008).

42. Park, C. Y. et al. Functional Knowledge Transfer for High-accuracy Prediction of Under-studied Biological Processes. PLoS Comput Biol **9**, e1002957 (2013).

43. Yamaguchi, M., Imai, F., Tonou-Fujimori, N. & Masai, I. Mutations in N-cadherin and a Stardust homolog, Nagie oko, affect cell-cycle exit in zebrafish retina. Mechanisms of Development **127**, 247–264 (2010).

44. Mancuso, C. A., Johnson, K. A., Liu, R. & Krishnan, A. Joint representation of molecular networks from multiple species improves gene classification. PLoS Comput Biol **20**, e1011773 (2024).

45. Barot, M., Gligorijević, V., Cho, K. & Bonneau, R. NetQuilt: deep multispecies network-based protein function prediction using homology-informed network similarity. Bioinformatics **37**, 2414–2422 (2021).

46. Katz, L. A New Status Index Derived from Sociometric Analysis. Psychometrika **18**, 39–43 (1953).

47. Liben-Nowell, D. & Kleinberg, J. The link prediction problem for social networks. in Proceedings of the twelfth international conference on Information and knowledge management 556–559 (ACM, 2003). doi:10.1145/956863.956972.

48. Gligorijevic, D. et al. Large-Scale Discovery of Disease-Disease and Disease-Gene Associations. Sci Rep **6**, (2016).

49. Sonawane, A. R. et al. Understanding Tissue-Specific Gene Regulation. Cell Reports **21**, 1077–1088 (2017).

50. Hekselman, I. & Yeger-Lotem, E. Mechanisms of tissue and cell-type specificity in heritable traits and diseases. Nat Rev Genet **21**, 137–150 (2020).

51. Greene, C. S. et al. Understanding multicellular function and disease with human tissue-specific networks. Nat Genet **47**, 569–576 (2015).

52. Cowen, L., Ideker, T., Raphael, B. J. & Sharan, R. Network propagation: a universal amplifier of genetic associations. Nat Rev Genet **18**, 551–562 (2017).

53. Picart-Armada, S. et al. Benchmarking network propagation methods for disease gene identification. PLoS Comput Biol **15**, e1007276 (2019).

54. Krishnan, A. et al. Genome-wide prediction and functional characterization of the genetic basis of autism spectrum disorder. Nat Neurosci **19**, 1454–1462 (2016).

55. Mastropietro, A., De Carlo, G. & Anagnostopoulos, A. XGDAG: explainable gene–disease associations via graph neural networks. Bioinformatics **39**, (2023).

56. Domazet-Loso, T. & Tautz, D. An Ancient Evolutionary Origin of Genes Associated with Human Genetic Diseases. Molecular Biology and Evolution **25**, 2699–2707 (2008).

57. Cai, J. J., Borenstein, E., Chen, R. & Petrov, D. A. Similarly Strong Purifying Selection Acts on Human Disease Genes of All Evolutionary Ages. Genome Biology and Evolution **1**, 131–144 (2009).

58. Lopez-Bigas, N. Genome-wide identification of genes likely to be involved in human genetic disease. Nucleic Acids Research **32**, 3108–3114 (2004).

59. Dickerson, J. E. & Robertson, D. L. On the Origins of Mendelian Disease Genes in Man: The Impact of Gene Duplication. Molecular Biology and Evolution **29**, 61–69 (2011).

60. Pellegrini, M., Marcotte, E. M., Thompson, M. J., Eisenberg, D. & Yeates, T. O. Assigning protein functions by comparative genome analysis: Protein phylogenetic profiles. Proc. Natl. Acad. Sci. U.S.A. **96**, 4285–4288 (1999).

61. Marcotte, E. M. et al. Detecting Protein Function and Protein-Protein Interactions from Genome Sequences. Science **285**, 751–753 (1999).

62. Maxwell, E. K. et al. Evolutionary profiling reveals the heterogeneous origins of classes of human disease genes: implications for modeling disease genetics in animals. BMC Evol Biol **14**, (2014).

63. Amberger, J. S., Bocchini, C. A., Scott, A. F. & Hamosh, A. OMIM.org: leveraging knowledge across phenotype–gene relationships. Nucleic Acids Research **47**, D1038–D1043 (2018).

64. Tabach, Y. et al. Human disease locus discovery and mapping to molecular pathways through phylogenetic profiling. Molecular Systems Biology **9**, (2013).

65. Dey, G., Jaimovich, A., Collins, Sean R., Seki, A. & Meyer, T. Systematic Discovery of Human Gene Function and Principles of Modular Organization through Phylogenetic Profiling. Cell Reports **10**, 993–1006 (2015).

66. Brown, S. D. M. et al. High-throughput mouse phenomics for characterizing mammalian gene function. Nat Rev Genet **19**, 357–370 (2018).

67. Peterson, K. A. & Murray, S. A. Progress towards completing the mutant mouse null resource. Mamm Genome **33**, 123–134 (2021).

68. Alliance of Genome Resources Consortium. Harmonizing model organism data in the Alliance of Genome Resources. Genetics **220**, iyac022 (2022).

69. Matentzoglu, N. et al. The Unified Phenotype Ontology : a framework for cross-species integrative phenomics. GENETICS **229**, (2025).

70. Hoehndorf, R., Schofield, P. N. & Gkoutos, G. V. PhenomeNET: a whole-phenome approach to disease gene discovery. Nucleic Acids Research **39**, e119–e119 (2011).

71. Smedley, D. et al. Next-generation diagnostics and disease-gene discovery with the Exomiser. Nat Protoc **10**, 2004–2015 (2015).

72. Alghamdi, S. M., Schofield, P. N. & Hoehndorf, R. Contribution of model organism phenotypes to the computational identification of human disease genes. Disease Models & Mechanisms **15**, dmm049441 (2022).

73. Althagafi, A., Zhapa-Camacho, F. & Hoehndorf, R. Prioritizing genomic variants through neuro-symbolic, knowledge-enhanced learning. Bioinformatics **40**, (2024).

74. Gherardini, P. F., Wass, M. N., Helmer-Citterich, M. & Sternberg, M. J. E. Convergent Evolution of Enzyme Active Sites Is not a Rare Phenomenon. Journal of Molecular Biology **372**, 817–845 (2007).

75. Tulipano, A., Donvito, G., Licciulli, F., Maggi, G. & Gisel, A. Gene analogue finder: a GRID solution for finding functionally analogous gene products. BMC Bioinformatics **8**, (2007).

76. Li, H. TreeFam: a curated database of phylogenetic trees of animal gene families. Nucleic Acids Research **34**, D572–D580 (2006).

77. Fan, J. et al. Functional protein representations from biological networks enable diverse cross-species inference. Nucleic Acids Research **47**, e51–e51 (2019).

78. Arsenescu, V. et al. MUNDO: protein function prediction embedded in a multispecies world. Bioinformatics Advances **2**, (2021).

79. Li, L. et al. Joint embedding of biological networks for cross-species functional alignment. Bioinformatics **39**, (2023).

80. Lin, Z. et al. Evolutionary-scale prediction of atomic-level protein structure with a language model. Science **379**, 1123–1130 (2023).

81. Rives, A. et al. Biological structure and function emerge from scaling unsupervised learning to 250 million protein sequences. Proc. Natl. Acad. Sci. U.S.A. **118**, (2021).

82. Hamamsy, T. et al. Protein remote homology detection and structural alignment using deep learning. Nat Biotechnol **42**, 975–985 (2023).

83. Hong, L. et al. Fast, sensitive detection of protein homologs using deep dense retrieval. Nat Biotechnol https://doi.org/10.1038/s41587-024-02353-6 (2024) doi:10.1038/s41587-024-02353-6.

84. Brixi, G. et al. Genome modeling and design across all domains of life with Evo 2. (2025) doi:10.1101/2025.02.18.638918.

85. McGary, K. L. et al. Systematic discovery of nonobvious human disease models through orthologous phenotypes. Proc. Natl. Acad. Sci. U.S.A. **107**, 6544–6549 (2010).

86. Lee, J., Shah, M., Ballouz, S., Crow, M. & Gillis, J. CoCoCoNet: conserved and comparative co-expression across a diverse set of species. Nucleic Acids Research **48**, W566–W571 (2020).

87.

El-Kebir, M. et al. xHeinz: an algorithm for mining cross-species network modules under a flexible conservation model. Bioinformatics **31**, 3147–3155 (2015).

88. Deshpande, R., Sharma, S., Verfaillie, C. M., Hu, W.-S. & Myers, C. L. A Scalable Approach for Discovering Conserved Active Subnetworks across Species. PLoS Comput Biol **6**, e1001028 (2010).

89. Zinman, G. E. et al. ModuleBlast: identifying activated sub-networks within and across species. Nucleic Acids Research **43**, e20–e20 (2014).

90. Zong, W. et al. Transcriptomic congruence analysis for evaluating model organisms. Proc. Natl. Acad. Sci. U.S.A. **120**, (2023).

91. Seok, J. et al. Genomic responses in mouse models poorly mimic human inflammatory diseases. Proc. Natl. Acad. Sci. U.S.A. **110**, 3507–3512 (2013).

92. Takao, K. & Miyakawa, T. Genomic responses in mouse models greatly mimic human inflammatory diseases. Proc. Natl. Acad. Sci. U.S.A. **112**, 1167–1172 (2014).

93. Putman, T. E. et al. The Monarch Initiative in 2024: an analytic platform integrating phenotypes, genes and diseases across species. Nucleic Acids Research **52**, D938–D949 (2023).

94. Normand, R. et al. Found In Translation: a machine learning model for mouse-to-human inference. Nat Methods **15**, 1067–1073 (2018).

95. Brubaker, D. K., Proctor, E. A., Haigis, K. M. & Lauffenburger, D. A. Computational translation of genomic responses from experimental model systems to humans. PLoS Comput Biol **15**, e1006286 (2019).

96. Brubaker, D. K. et al. An interspecies translation model implicates integrin signaling in infliximab-resistant inflammatory bowel disease. Sci. Signal. **13**, (2020).

97. Suarez-Lopez, L. et al. Cross-species transcriptomic signatures predict response to MK2 inhibition in mouse models of chronic inflammation. iScience **24**, 103406 (2021).

98. Lee, M. J. et al. Computational Interspecies Translation Between Alzheimer's Disease Mouse Models and Human Subjects Identifies Innate Immune Complement, TYROBP, and TAM Receptor Agonist Signatures, Distinct From Influences of Aging. Front. Neurosci. **15**, (2021).

99. Carroll, M. J., Garcia-Reyero, N., Perkins, E. J. & Lauffenburger, D. A. Translatable pathways classification (TransPath-C) for inferring processes germane to human biology from animal studies data: example application in neurobiology. Integrative Biology **13**, 237–245 (2021).

100. Russkikh, N. et al. Style transfer with variational autoencoders is a promising approach to RNA-Seq data harmonization and analysis. Bioinformatics **36**, 5076–5085 (2020).

101. Lotfollahi, M., Wolf, F. A. & Theis, F. J. scGen predicts single-cell perturbation responses. Nat Methods **16**, 715–721 (2019).

102. Meimetis, N. et al. AutoTransOP: translating omics signatures without orthologue requirements using deep learning. npj Syst Biol Appl **10**, (2024).

103. Subramanian, A. et al. Gene set enrichment analysis: A knowledge-based approach for interpreting genome-wide expression profiles. Proc. Natl. Acad. Sci. U.S.A. **102**, 15545–15550 (2005).

104. Cai, M., Hao Nguyen, C., Mamitsuka, H. & Li, L. XGSEA: CROSS-species gene set enrichment analysis via domain adaptation. Briefings in Bioinformatics **22**, (2021).

105. Alié, A. et al. The ancestral gene repertoire of animal stem cells. Proc. Natl. Acad. Sci. U.S.A. **112**, (2015).

106. Shafer, M. E. R. Cross-Species Analysis of Single-Cell Transcriptomic Data. Front. Cell Dev. Biol. **7**, (2019).

107. Luecken, M. D. et al. Benchmarking atlas-level data integration in single-cell genomics. Nat Methods **19**, 41–50 (2021).

108. Butler, A., Hoffman, P., Smibert, P., Papalexi, E. & Satija, R. Integrating single-cell transcriptomic data across different conditions, technologies, and species. Nat Biotechnol **36**, 411–420 (2018).

109. Welch, J. D. et al. Single-Cell Multi-omic Integration Compares and Contrasts Features of Brain Cell Identity. Cell **177**, 1873–1887.e17 (2019).

110. Bailon-Zambrano, R. et al. Variable paralog expression underlies phenotype variation. eLife **11**, (2022).

111. Forslund, K., Schreiber, F., Thanintorn, N. & Sonnhammer, E. L. L. OrthoDisease: tracking disease gene orthologs across 100 species. Briefings in Bioinformatics **12**, 463–473 (2011).

112. Song, Y., Miao, Z., Brazma, A. & Papatheodorou, I. Benchmarking strategies for cross-species integration of single-cell RNA sequencing data. Nat Commun **14**, (2023).

113. Tarashansky, A. J. et al. Mapping single-cell atlases throughout Metazoa unravels cell type evolution. eLife **10**, (2021).

114. Rosen, Y. et al. Toward universal cell embeddings: integrating single-cell RNA-seq datasets across species with SATURN. Nat Methods **21**, 1492–1500 (2024).

115. Theodoris, C. V. et al. Transfer learning enables predictions in network biology. Nature **618**, 616–624 (2023).

116. Cui, H. et al. scGPT: toward building a foundation model for single-cell multi-omics using generative AI. Nat Methods **21**, 1470–1480 (2024).

117. Galkin, F. et al. Precious3GPT: Multimodal Multi-Species Multi-Omics Multi-Tissue Transformer for Aging Research and Drug Discovery. (2024) doi:10.1101/2024.07.25.605062.

118. Yang, X. et al. GeneCompass: deciphering universal gene regulatory mechanisms with a knowledge-informed cross-species foundation model. Cell Res **34**, 830–845 (2024).

119. Rosen, Y. et al. Universal Cell Embeddings: A Foundation Model for Cell Biology. (2023) doi:10.1101/2023.11.28.568918.

120. Marian, A. J. Modeling Human Disease Phenotype in Model Organisms. Circulation Research **109**, 356–359 (2011).

121. Li, Y. & Agarwal, P. A Pathway-Based View of Human Diseases and Disease Relationships. PLoS ONE **4**, e4346 (2009).

122. Cui, P., Ma, X., Li, H., Lang, W. & Hao, J. Shared Biological Pathways Between Alzheimer's Disease and Ischemic Stroke. Front. Neurosci. **12**, (2018).

123. Gokuladhas, S., Schierding, W., Golovina, E., Fadason, T. & O'Sullivan, J. Unravelling the Shared Genetic Mechanisms Underlying 18 Autoimmune Diseases Using a Systems Approach. Front. Immunol. **12**, (2021).

124. Tokarek, J. et al. Molecular Processes Involved in the Shared Pathways between Cardiovascular Diseases and Diabetes. Biomedicines **11**, 2611 (2023).

125. Stoney, R., Robertson, D. L., Nenadic, G. & Schwartz, J.-M. Mapping biological process relationships and disease perturbations within a pathway network. npj Syst Biol Appl **4**, (2018).

126. Hickey, S. L., McKim, A., Mancuso, C. A. & Krishnan, A. A network-based approach for isolating the chronic inflammation gene signatures underlying complex diseases towards finding new treatment opportunities. Front. Pharmacol. **13**, (2022).

127. Buphamalai, P., Kokotovic, T., Nagy, V. & Menche, J. Network analysis reveals rare disease signatures across multiple levels of biological organization. Nat Commun **12**, (2021).

128. Gargano, M. A. et al. The Human Phenotype Ontology in 2024: phenotypes around the world. Nucleic Acids Research **52**, D1333–D1346 (2023).

129. Blair, D. R., Hoffmann, T. J. & Shieh, J. T. Common genetic variation associated with Mendelian disease severity revealed through cryptic phenotype analysis. Nat Commun **13**, (2022).

130. Maassen, W. et al. Curation and expansion of the Human Phenotype Ontology for systemic autoinflammatory diseases improves phenotype-driven disease-matching. Front. Immunol. **14**, (2023).

131. Brunet, F. G. et al. Gene Loss and Evolutionary Rates Following Whole-Genome Duplication in Teleost Fishes. Molecular Biology and Evolution **23**, 1808–1816 (2006).

132. True, J. R. & Haag, E. S. Developmental system drift and flexibility in evolutionary trajectories. Evolution and Development **3**, 109–119 (2001).

133. Ghadie, M. A., Coulombe-Huntington, J. & Xia, Y. Interactome evolution: insights from genome-wide analyses of protein–protein interactions. Current Opinion in Structural Biology **50**, 42–48 (2018).

134. Ecovoiu, A. Al., Ratiu, A. C., Micheu, M. M. & Chifiriuc, M. C. Inter-Species Rescue of Mutant Phenotype—The Standard for Genetic Analysis of Human Genetic Disorders in Drosophila melanogaster Model. IJMS **23**, 2613 (2022).

135. Kachroo, A. H. et al. Systematic humanization of yeast genes reveals conserved functions and genetic modularity. Science **348**, 921–925 (2015).

136. Futuyma, D. J. Evolution. (Sinauer Associates, Inc. Publishers, Sunderland, Massachusetts U.S.A, 2013).

137. Szklarczyk, D. et al. The STRING database in 2023: protein–protein association networks and functional enrichment analyses for any sequenced genome of interest. Nucleic Acids Research **51**, D638–D646 (2022).

138. Szymborski, J. & Emad, A. INTREPPPID - An Orthologue-Informed Quintuplet Network for Cross-Species Prediction of Protein-Protein Interaction. (2024) doi:10.1101/2024.02.13.580150.

139.

Walhout, A. J. M. et al. Protein Interaction Mapping in <i>C. elegans</i> Using Proteins Involved in Vulval Development. Science **287**, 116–122 (2000).

140. Yu, H. et al. Annotation Transfer Between Genomes: Protein–Protein Interologs and Protein–DNA Regulogs. Genome Res. **14**, 1107–1118 (2004).

141. Wong, A. K., Krishnan, A., Yao, V., Tadych, A. & Troyanskaya, O. G. IMP 2.0: a multi-species functional genomics portal for integration, visualization and prediction of protein functions and networks. Nucleic Acids Res **43**, W128–W133 (2015).

142. Sealfon, R. S. G., Wong, A. K. & Troyanskaya, O. G. Machine learning methods to model multicellular complexity and tissue specificity. Nat Rev Mater **6**, 717–729 (2021).

143. Saha, A. et al. Co-expression networks reveal the tissue-specific regulation of transcription and splicing. Genome Res. **27**, 1843–1858 (2017).

144. Burns, J. J. R. et al. Addressing noise in co-expression network construction. Briefings in Bioinformatics **23**, (2021).

145. Xue, S. et al. Applying differential network analysis to longitudinal gene expression in response to perturbations. Front. Genet. **13**, (2022).

146. Liu, R., Yuan, H., Johnson, K. & Krishnan, A. CONE: COntext-specific Network Embedding via Contextualized Graph Attention. Proceedings of the 19th Machine Learning in Computational Biology meeting **261**, 53–71 (2024).

147. Li, M. M. et al. Contextual AI models for single-cell protein biology. Nat Methods **21**, 1546–1557 (2024).

148. Zitnik, M. & Leskovec, J. Predicting multicellular function through multi-layer tissue networks. Bioinformatics **33**, i190–i198 (2017).

149. Evans, R. et al. Protein complex prediction with AlphaFold-Multimer. (2021) doi:10.1101/2021.10.04.463034.

150. Zhou, J. et al. Deep learning sequence-based ab initio prediction of variant effects on expression and disease risk. Nat Genet **50**, 1171–1179 (2018).

151. Sokolova, K. et al. Atlas of primary cell-type-specific sequence models of gene expression and variant effects. Cell Reports Methods **3**, 100580 (2023).

152. Rhie, A. et al. Towards complete and error-free genome assemblies of all vertebrate species. Nature **592**, 737–746 (2021).

153. Li, M. M., Huang, K. & Zitnik, M. Graph representation learning in biomedicine and healthcare. Nat. Biomed. Eng **6**, 1353–1369 (2022).

154. Bradshaw, M. S., Gaskell, A. & Layer, R. M. The effects of biological knowledge graph topology on embedding-based link prediction. (2024) doi:10.1101/2024.06.10.598277.

155. O'Neil, S. T. et al. Phenomics Assistant: An Interface for LLM-based Biomedical Knowledge Graph Exploration. (2024) doi:10.1101/2024.01.31.578275.

156. Gaudet, P. & Dessimoz, C. Gene Ontology: Pitfalls, Biases, and Remedies. in Methods in Molecular Biology 189–205 (Springer New York, 2016). doi:10.1007/978-1-4939-3743-1_14.

157. Caufield, J. H. et al. Structured prompt interrogation and recursive extraction of semantics (SPIRES): A method for populating knowledge bases using zero-shot learning. arXiv https://doi.org/10.48550/arxiv.2304.02711 (2023) doi:10.48550/arxiv.2304.02711.

158. Howe, K. et al. The zebrafish reference genome sequence and its relationship to the human genome. Nature **496**, 498–503 (2013).

159. Russell, J. J. et al. Non-model model organisms. BMC Biol **15**, (2017).

160. Smith, J. J. et al. A chromosome-scale assembly of the axolotl genome. Genome Res. **29**, 317–324 (2019).

161. Kourakis, M. J. & Smith, W. C. An organismal perspective on C. intestinalis development, origins and diversification. eLife **4**, (2015).